\documentclass[12pt]{iopart}
\usepackage{epsfig}
\usepackage{graphicx}
\begin{document}
\title[]{Weak value amplification of atomic cat states}
\author{Sumei Huang$^{1,2}$, Girish S. Agarwal$^2$}
\address{$^1$Department of Electrical and Computer Engineering, National University of
Singapore, 4 Engineering Drive 3, Singapore 117583}
\address{$^2$Department of Physics, Oklahoma State University, Stillwater, Oklahoma 74078, USA}
\eads{\mailto{girish.agarwal@okstate.edu}}
\date{\today}
\begin{abstract}
We show the utility of the weak value amplification to observe the quantum interference between two close lying atomic coherent states in a post-selected atomic cat state, produced in a system of $N$ identical two-level atoms weakly interacting with a single photon field. Through the observation of the negative parts of the Wigner distribution of the post-selected atomic cat state, we find that the post-selected atomic cat state becomes more nonclassical when the post-selected polarization state of the single photon field tends toward becoming orthogonal to its pre-selected state. We show that the small phase shift in the post-selected atomic cat state can be amplified via measuring the peak shift of its phase distribution when the post-selected state of the single photon field is nearly orthogonal to its pre-selected state. We find that the amplification factor of 15 [5] can be obtained for a sample of 10 [100] atoms. This effectively provides us with a method to discriminate two close lying states on the Bloch sphere. We discuss possible experimental implementation of the scheme, and conclude with a discussion of the Fisher information.\\

\end{abstract}

\maketitle

\section{Introduction}
The weak value amplification of the observables is finding increasing number of
applications in the study of a variety of physical systems \cite{Aharonov,Duck,Leggett,BoydRMP,Nori}. Although originally
formulated for quantum systems, many past and current applications include
applications to classical light beams. For example the first observation of the weak
value amplification was in the context of a Gaussian beam propagating through a
birefringent medium \cite{Hulet}. Other important applications of weak value amplification
include observation of spin Hall effect of light \cite{Hosten},
Goos-H\"{a}nchen shifts and various generalizations \cite{Jayaswal1,Jayaswal2,Goswami}, angular shifts of light beams \cite{Boyd},
enhancement of interferometric results \cite{Torres,Brunner,Bruder}. Weak value amplification has been used to measure the state of polarization of light beam on Poincare sphere using optical vortex beams \cite{Shikano}. It is intriguing that a
concept formulated for quantum systems has so many profound applications in
the context of classical light beams. Aiello showed in a formal way how weak value amplification works for beams of light \cite{Aiello}. Lundeen
and coworkers used weak value amplification to get the wavefront of a single
photon \cite{Lundeen}. Steinberg \cite{Hayat} proposed the applications in
the measurement of interaction between two Fermions. Weak value amplification has been proposed to measure the presence of an additional charge in Ahranov Bohm interferometer \cite{Gefen}. Pryde et. al. experimentally determined weak values for a single photon's polarization via a weak value amplification \cite{Pryde}. Starling et. al. used the weak value amplification to enhance frequency shift resolution in a Sagnac interferometer \cite{Starling}. While most examine the amplification of the small shifts, several have examined the question of improvement in sensitivity \cite{Jordan,Zhang13,Caves,Yamamoto,Jordanprx,Jordanexp,Alves,Bagan,KneeA,KneeX,FerrieL,KneeArxiv,Gross} of the measurement. The weak value amplification can worsen the metrological performance \cite{KneeArxiv} for example the technical noise or decoherence can not be completely overcome \cite{KneeA,KneeX} although a significant improvement can be obtained \cite{Jordanprx}. An optimal strategy would be to retain full data \cite{Bruder,FerrieL} and do a weak measurement. However advantages of this technique are not exclusive to this technique \cite{Gross}.

In this paper we show the great advantage offered by weak value amplification for
studying quantum mechanical cat states for atoms. The cat states are the linear
superposition of two coherent states on the Bloch sphere
\begin{equation}\label{1}
|\Psi_{cat}\rangle= a|\theta_1,\phi_1\rangle + b|\theta_2,\phi_2\rangle,
\end{equation}
where the coefficients $a$ and $b$ represent the probability amplitudes for the atomic system to be in the atomic coherent states $|\theta_1,\phi_1\rangle$ and $|\theta_2,\phi_2\rangle$, respectively.
The quantum interferences in cat state are most prominent if the two coherent
states are close on the Bloch sphere \cite{Gerry,Agarwal97}. The study of quantum interferences is
greatly aided by the weak value amplification otherwise these are difficult to
observe. The weak value amplification gives us the capability to resolve two close
lying coherent states. We look at the interaction of a single photon with an
ensemble of atoms prepared in a coherent state \cite{McConnell,Agarwal}. The interaction produces an
entangled state of the photon polarization variables with the coherent states of
the atomic ensemble. We use preselection and postselection of the polarization
states of the photon. The postselected polarization is nearly orthogonal to the
input polarization. This enables us to magnify the weak values associated with the
measurements of the phase $\varphi$. Although in our work we produce cat states by heralding i.e. by detection of a photon, there are many methods to produce cat states \cite{Agarwal97,Haroche,Brune,Gerry98,Raimond,Simon,TO,Dooley,Rao}. The most prominent method is to use atomic systems dispersively interacting with a cavity field \cite{Agarwal97,Haroche,Brune,Gerry98,Raimond}. The importance of cat states in quantum optics hardly needs to be emphasized as these have important nonclassical properties \cite{Gerry,Agarwal97} and are important in precision measurements \cite{McConnell,Leibfried}.

The organization of this paper is as follows: In section 2, we
introduce the model of the interacting atom-field system. In section 3,
we make a weak value amplification on the atom-field system so that the post-selected atomic cat state is generated. In section 4, we present the variation of the Wigner distribution of the post-selected atomic cat state when the overlap of the initial and final states of the field changes. In section 5, we
show that the small phase shift in the post-selected atomic cat state can be amplified by choosing nearly orthogonal pre-selection and post-selection of the single photon field. In this section we also discuss how the state tomography of the post-selected cat state can be done. In section 6, we discuss the weak value amplification for our atomic cat states using the quantum Fisher information and show that
the Fisher information in the meter and the classical Fisher information yields the quantum Fisher information of the full meter system state. This is in agreement with recent conclusions in several papers \cite{Jordanprx,Alves}. We conclude our paper in the final section.

\section{Atomic systems and the effective interaction Hamiltonian}
\begin{figure}[htp]
\begin{center}
\scalebox{0.9}{\includegraphics{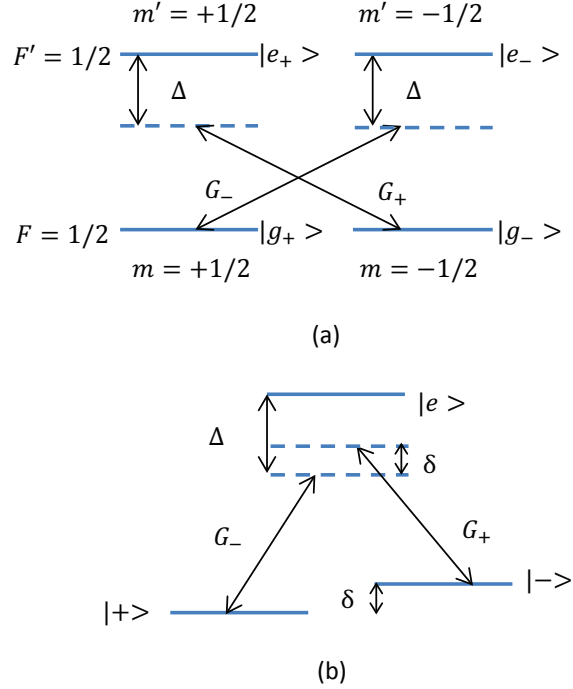}}
\caption{\label{Fig1} (Color online) (a) Schematic diagram of an atomic ensemble coupled to two polarized modes of a single photon field. (b) Atoms with two ground states $|+\rangle$ and $|-\rangle$ are coupled to an excited state $|e\rangle$ via two degenerate polarized modes.}
\end{center}
\end{figure}

We consider an ensemble of $N$ identical two-level atoms interacting with two orthogonally polarized modes of a single photon field with frequency $\omega_{f}$ denoted by creation (annihilation) operators $a_{-}^{\dag}$, $a_{+}^{\dag}$ ($a_{-}$, $a_{+}$) as shown in Fig.~\ref{Fig1}(a). The two-level atoms have degenerate ground states $|g_{\pm}\rangle$ and excited states $|e_{\pm}\rangle$, separated by an energy of $\hbar\omega_0$. According to the angular-momentum selection rules, the transitions $|g_{+}\rangle\leftrightarrow|e_{+}\rangle$ and $|g_{-}\rangle\leftrightarrow|e_{-}\rangle$ are forbidden, only the transitions $|g_{+}\rangle\leftrightarrow|e_{-}\rangle$ and $|g_{-}\rangle\leftrightarrow|e_{+}\rangle$ are allowed. Moreover, the levels $|g_{+}\rangle$ and $|e_{-}\rangle$ are coupled by the field mode $a_{-}$, and the levels $|g_{-}\rangle$ and $|e_{+}\rangle$ are coupled by the field mode $a_{+}$. Their coupling strengthes are $G_{-}$ and $G_{+}$, respectively. The Hamiltonian of the combined system of the atoms and the field \cite{Agarwal} takes the form

\begin{eqnarray}\label{2}
H&=&\hbar\frac{\omega_0}{2}\sum_{i=1}^{N}(|e_{-}\rangle\langle e_{-}|-|g_{+}\rangle\langle g_{+}|)_{i}
+\hbar\frac{\omega_0}{2}\sum_{i=1}^{N}(|e_{+}\rangle\langle e_{+}|-|g_{-}\rangle\langle g_{-}|)_{i}\nonumber\\
& &+\hbar\omega_{f}(a_{+}^{\dag}a_{+}+a_{-}^{\dag}a_{-})+\hbar\left(G_{-}a_{-}\sum_{i=1}^{N}|e_{-}\rangle\langle g_{+}|_{i}+h.c.\right)\nonumber\\
& &+\hbar\left(G_{+}a_{+}\sum_{i=1}^{N}|e_{+}\rangle\langle g_{-}|_{i}+h.c.\right).
\end{eqnarray}
In Eq. (\ref{2}), the first two terms are the atomic excitation Hamiltonian, the third term is the free field Hamiltonian, the last two terms are the atom-field interaction Hamiltonian.
In the dispersive limit in which the detuning between the atomic transitions and the field modes is much
larger than the atom-field coupling strengthes i.e., $\omega_{0}-\omega_{f}=\Delta\gg\{|G_{+}|,|G_{-}|\}$, the Hamiltonian equation (\ref{2}) can be reduced to
\begin{eqnarray}\label{3}
H_{eff}&=&\hbar\frac{|G_{-}|^2}{\Delta}a_{-}^{\dag}a_{-}\sum_{i=1}^{N}(|e_{-}\rangle\langle e_{-}|-|g_{+}\rangle\langle g_{+}|)_{i}\nonumber\\& &+\hbar\frac{|G_{+}|^2}{\Delta}a_{+}^{\dag}a_{+}\sum_{i=1}^{N}(|e_{+}\rangle\langle e_{+}|-|g_{-}\rangle\langle g_{-}|)_{i}\nonumber\\
& &+\frac{2\hbar}{\Delta}\left(|G_{-}|^2\sum_{i=1}^{N}|e_{-}\rangle\langle e_{-}|_{i}+|G_{+}|^2\sum_{i=1}^{N}|e_{+}\rangle\langle e_{+}|_{i}\right).
\end{eqnarray}
Note that there are no couplings between the ground states $|g_{\pm}\rangle_{i}$ ($i=1,\cdots, N$) and the excited ones.
For simplicity, we assume that the coupling strengths have identical amplitudes $|G_{-}|=|G_{+}|=G$. If one starts from the initial ground state, the relevant interaction Hamiltonian between the atoms and the field modes is written as
\begin{equation}\label{4}
H_{int}=\hbar \phi_{0}N_{z}J_{z},
\end{equation}
where
\begin{equation}\label{5}
N_{z}=a_{+}^{\dag}a_{+}-a_{-}^{\dag}a_{-}
\end{equation}
is the field operator,
\begin{equation}\label{6}
J_{z}=\frac{1}{2}\sum_{i=1}^{N}(|g_{+}\rangle\langle g_{+}|-|g_{-}\rangle\langle g_{-}|)_{i}
\end{equation}
is the collective atomic operator,
 and $\phi_{0}=G^2/\Delta$ is the coupling constant between $N_{z}$ and $J_{z}$.

Another important example of optical transitions leading to the interaction Hamiltonian (\ref{4}) would be three level atoms \cite{McConnell} with ground levels $|\pm\rangle$ coupled to an excited state $|e\rangle$ by a far off resonant field, as shown in Fig.~\ref{Fig1}(b). Under the assumption that the frequency separation between $|\pm\rangle$ is such that the two photon Raman coupling between $|+\rangle$ and $|-\rangle$ is negligible, the effective interaction between the atoms and the field is described by Eq. (\ref{4}).

In what follows we assume that the initial state of the atomic sample is an atomic coherent state $|\theta,\phi\rangle$ with
a definite angular momentum value $j$, which can be created by rotating the ground state $|j,-j\rangle$ on a
Bloch sphere by an angle $\theta$ around an axis defined as $\vec{n}=(-\sin\phi,\cos\phi,0)$ \cite{Arecchi}
\begin{eqnarray}\label{7}
|\theta,\phi\rangle&=&R_{\theta,\phi}|j,-j\rangle=\exp(i\theta\vec{n}\cdot\vec{J})|j,-j\rangle\nonumber\\
&=&\exp(\zeta J_{+}-\zeta^{*} J_{-})|j,-j\rangle,
\end{eqnarray}
where $\zeta=\frac{\theta}{2}e^{-i\phi}$, $0\leq\theta\leq\pi$, $0\leq\phi\leq2\pi$, $J_{-}=\sum_{i=1}^{N}(|g_{-}\rangle\langle g_{+}|)_{i}$, $J_{+}=\sum_{i=1}^{N}(|g_{+}\rangle\langle g_{-}|)_{i}$, and $|j,-j\rangle=\Pi_{i=1}^{N}|g_{-}\rangle_{i}$ indicates a state for which all atoms are in the ground state. The atomic coherent state also can be expressed as \cite{Arecchi}
\begin{eqnarray}\label{8}
|\theta,\phi\rangle&=&\sum_{m=-j}^{+j}\left(\begin{array}{c}
2j\\
j+m\\
\end{array}\right)^{\frac{1}{2}}\sin^{j+m}\left(\frac{\theta}{2}\right)\cos^{j-m}\left(\frac{\theta}{2}\right) e^{-i(j+m)\phi}|j,m\rangle,
\end{eqnarray}
where $|j,m\rangle$ is the simultaneous eigenstate of $J^2$ and $J_{z}$,
\begin{eqnarray}\label{9}
J^2|j,m\rangle=j(j+1)|j,m\rangle,\nonumber\\
J_{z}|j,m\rangle=m|j,m\rangle,
\end{eqnarray}
where $j=\frac{N}{2}$, $m=-j,-j+1,\cdots,j-1,j$, and $N$ is a total number of atoms in the sample.
The atomic coherent states are in general not orthogonal except for antipodal points. The modulus squared of the inner product of two atomic coherent states is
\begin{equation}\label{10}
|\langle \theta,\phi|\theta',\phi'\rangle|^2=\left(\cos^2\frac{\Theta}{2}\right)^{2j},
\end{equation}
where $\Theta$ is the angle
between the directions $(\theta, \phi)$ and $(\theta', \phi')$ on the Bloch sphere, and $\cos \Theta=\cos\theta\cos\theta'+\sin\theta\sin\theta'\cos(\phi-\phi')$. For the special cases ($\theta'-\theta=\pm\pi$, $\phi'=\phi$; or $\theta'=\theta=\pi/2$, $\phi-\phi'=\pm\pi$), $\cos \Theta=-1$, $\Theta=\pi$, $|\langle \theta,\phi|\theta',\phi'\rangle|^2=0$, the two atomic coherent states are orthogonal. The atomic coherent states can be produced by classical driving fields of constant amplitude \cite{Reca1, Reca2}.

We assume that the initial state of the single photon field is given by
\begin{equation}\label{11}
|\Psi_{f}\rangle=c_{+}|1_{+},0_{-}\rangle+c_{-}|0_{+},1_{-}\rangle,
\end{equation}
where the coefficients $c_{+}$ and $c_{-}$ are the probability amplitudes to find the light in the states $|1_{+},0_{-}\rangle$ and $|0_{+},1_{-}\rangle$, respectively, and $|c_{+}|^2+|c_{-}|^2=1$. Note that the $|\Psi_{f}\rangle$ corresponds to an elliptically polarized light. When $c_{+}=0$ or $c_{-}=0$, the $|\Psi_{f}\rangle$ is circularly polarized. When $c_{+}=c_{-}$, the light is $x$ polarized. Note that single photons can be heralded \cite{Brida,Lvovsky} or a single nitrogen-vacancy color center in a diamond nanocrystal \cite{All, Beve}.
Thus the initial state of atom-field system is $|\theta,\phi\rangle|\Psi_{f}\rangle$, which is a product state of the two subsystems of the field and the atomic ensemble.

\section{Weak interaction of the atomic ensemble with the single photon field and the post-selected atomic state}
We next study the evolution of the system under the interaction (\ref{4}). This is given by
\begin{eqnarray}\label{12}
|\Psi_{at-f}\rangle&=&\exp\{-i\frac{H_{int}}{\hbar}t\}|\theta,\phi\rangle|\Psi_{f}\rangle,\nonumber\\
&=&c_{+}|\theta,\phi+\Omega\rangle|1_{+},0_{-}\rangle\exp(ij\Omega)+c_{-}|\theta,\phi-\Omega\rangle|0_{+},1_{-}\rangle\exp(-ij\Omega),\nonumber\\
\end{eqnarray}
where $t$ is the atom-field interaction time. Expression (\ref{12}) indicates that the atoms are entangled with the single photon field due to the atom-field coupling. The two coherent states in (\ref{12}) differ in phase by $2\Omega$, $\Omega=\phi_{0}t$. The interaction of the single photon with the medium is expected to be weak and hence $\Omega$ is small. Thus the question of distinguishing between two such coherent states arises. It is here that the weak value amplification offers great advantage. We make a preselection and postselection of the states of the polarization of the single photon. For simplicity we assume that initially the field is linearly polarized with $c_{+}=c_{-}=\frac{1}{\sqrt{2}}$ i.e., $|\Psi_{f}\rangle=\frac{1}{\sqrt{2}}(|1_{+},0_{-}\rangle+|0_{+},1_{-}\rangle)$. We take the post-selected polarization state as
 \begin{equation}\label{13}
 |\Psi_{ph}\rangle=\sin(\gamma-\frac{\pi}{4})|1_{+},0_{-}\rangle+\cos(\gamma-\frac{\pi}{4})|0_{+},1_{-}\rangle,
  \end{equation}
  in which $\gamma$ is a small angle, and is controlled by the polarizer. Hence the overlap between the pre-selected and post-selected states of the field is \begin{equation}\label{14}
  |\langle \Psi_{ph}|\Psi_{f}\rangle|^2=\sin^2\gamma,
   \end{equation}
   which depends on the parameter $\gamma$. When $\gamma=0$, $\langle \Psi_{ph}|\Psi_{f}\rangle=0$, the pre-selected and post-selected states of the field are orthogonal. For small $\gamma$, the pre and post-selected states are nearly orthogonal. After the weak interaction between the atoms and the field, we project the state of the system $|\Psi_{at-f}\rangle$ onto a final state of the single photon field $|\Psi_{ph}\rangle$, we obtain the normalized atomic Schr\"{o}dinger cat state
\begin{eqnarray}\label{15}
|\Psi_{cat}\rangle&=&\frac{1}{\mathcal{N}}\langle \Psi_{ph}|\Psi_{at-f}\rangle\nonumber\\
&=&\frac{1}{\mathcal{N}}\Big[\frac{1}{\sqrt{2}}\sin(\gamma-\frac{\pi}{4})\exp(ij\Omega)|\theta,\phi+\Omega\rangle\nonumber\\& &+\frac{1}{\sqrt{2}}\cos(\gamma-\frac{\pi}{4})\exp(-ij\Omega)|\theta,\phi-\Omega\rangle\Big],
\end{eqnarray}
which is a superposition of two distinct atomic coherent states $|\theta,\phi+\Omega\rangle$ and $|\theta,\phi-\Omega\rangle$ rotating in opposite directions in phase space. Thus the weak interaction of the atoms with the single photon field, followed by the state-selective measurement on the single photon field, has transformed the initial atomic coherent state $|\theta,\phi\rangle$ into a post-selected atomic cat state with two components $|\theta,\phi+\Omega\rangle$ and $|\theta,\phi-\Omega\rangle$ with small opposite phase shifts $\Omega$. Here $\mathcal{N}$ is the normalization factor given by
\begin{eqnarray}\label{16}
\mathcal{N}^2&=&|\langle \Psi_{ph}|\Psi_{at-f}\rangle|^{2}\nonumber\\
&=&\frac{1}{2}\Bigg[\sum^{+j}_{m=-j}\left(\begin{array}{c} 2j \\ j+m\end{array}\right)\sin^{2(j+m)}\left(\frac{\theta}{2}\right)\cos^{2(j-m)}\left(\frac{\theta}{2}\right)\nonumber\\
& &-\cos(2\gamma)\sum^{+j}_{m=-j}\left(\begin{array}{c} 2j \\ j+m\end{array}\right)\sin^{2(j+m)}\left(\frac{\theta}{2}\right)\cos^{2(j-m)}\left(\frac{\theta}{2}\right)\cos(2m\Omega)\Bigg].\nonumber\\
\end{eqnarray}
In what follows we study the post-selected state for $\theta=\pi/2$ and $\phi=0$. This lies in the equatorial plane and has maximum coherence. In spin language this means that the spin is $x$ polarized at $t=0$.

\section{The Wigner function of the post-selected atomic cat state}
In this section, we use the Wigner function to quantify the nonclassicality of the post-selected atomic cat state. The Wigner function of the atomic cat state is defined as \cite{Agarwalbook}
\begin{equation}\label{17}
W(\alpha,\beta)=\sqrt{\frac{2j+1}{4\pi}}\sum_{KQ}\langle \Psi_{cat}|T_{KQ}^{\dag}|\Psi_{cat}\rangle Y_{KQ}(\alpha,\beta),
\end{equation}
where
\begin{eqnarray}\label{18}
T_{KQ}&=&\sum_{m_{1},m_{2}}(-1)^{j-m_{1}}(2K+1)^{\frac{1}{2}}\left(\begin{array}{ccc}j & K &j\\
-m_{1}& Q& m_{2}\end{array}\right)|j,m_{1}\rangle\langle j,m_{2}|
\end{eqnarray}
is the state-multipole operators \cite{Agarwal81}, $\left(\begin{array}{ccc}j & K &j\\
-m_{1}& Q& m_{2}\end{array}\right)$ is the Wigner $3j$ symbol, $K$ is an integer taking values 0, 1, 2, $\cdots$, $2j$, $-K\leq Q\leq K$, and
$Y_{KQ}(\alpha,\beta)$ is the spherical harmonic. It can be interpreted as a quasiprobability distribution in phase space. It satisfies the normalization condition
\begin{equation}\label{19}
\int W(\alpha,\beta)\sin\alpha\ \mbox{d}\alpha\ \mbox{d}\beta=1.
\end{equation}

After some calculations, we obtain the following result for the Wigner function
\begin{eqnarray}\label{20}
& &W(\alpha,\beta)=\frac{1}{\mathcal{N}^2}\left(\frac{1}{2}\right)^{2j+1}\sqrt{\frac{2j+1}{4\pi}}\nonumber\\
& &\times\Bigg\{\sin^2(\gamma-\frac{\pi}{4})\sum_{KQ}\sum_{m'=-j}^{+j}
\left(\begin{array}{c}2j\\j+m'\\\end{array}\right)^{\frac{1}{2}}e^{iQ\Omega}(-1)^{j-m'}(2K+1)^{\frac{1}{2}}\nonumber\\& &\times\left(\begin{array}{ccc}j & K &j\\
-m'& Q& m'-Q\end{array}\right)^{*}\left(\begin{array}{c}2j\\j+m'-Q\\\end{array}\right)^{\frac{1}{2}}Y_{KQ}(\alpha,\beta)\nonumber\\
& &+\cos^2(\gamma-\frac{\pi}{4})\sum_{KQ}\sum_{m'=-j}^{+j}
\left(\begin{array}{c}2j\\j+m'\\\end{array}\right)^{\frac{1}{2}}e^{-iQ\Omega}(-1)^{j-m'}(2K+1)^{\frac{1}{2}}\nonumber\\& &\times\left(\begin{array}{ccc}j & K &j\\
-m'& Q& m'-Q\end{array}\right)^{*}\left(\begin{array}{c}2j\\j+m'-Q\\\end{array}\right)^{\frac{1}{2}}Y_{KQ}(\alpha,\beta)\nonumber\\
& &-\cos(2\gamma)\sum_{KQ}\sum_{m'=-j}^{+j}
\left(\begin{array}{c}2j\\j+m'\\\end{array}\right)^{\frac{1}{2}}\cos[(2m'-Q)\Omega](-1)^{j-m'}(2K+1)^{\frac{1}{2}}\nonumber\\& &\times\left(\begin{array}{ccc}j & K &j\\
-m'& Q& m'-Q\end{array}\right)^{*}\left(\begin{array}{c}2j\\j+m'-Q\\\end{array}\right)^{\frac{1}{2}}Y_{KQ}(\alpha,\beta)\Bigg\},
\end{eqnarray}
in which the first term and the second term correspond to the contributions of the individual coherent states $|\frac{\pi}{2},\Omega\rangle$ and $|\frac{\pi}{2},-\Omega\rangle$ in the atomic cat state, respectively, the last term represents the interference between the two atomic coherent states.

\begin{figure}[htp]
\begin{center}
\scalebox{0.3}{\includegraphics{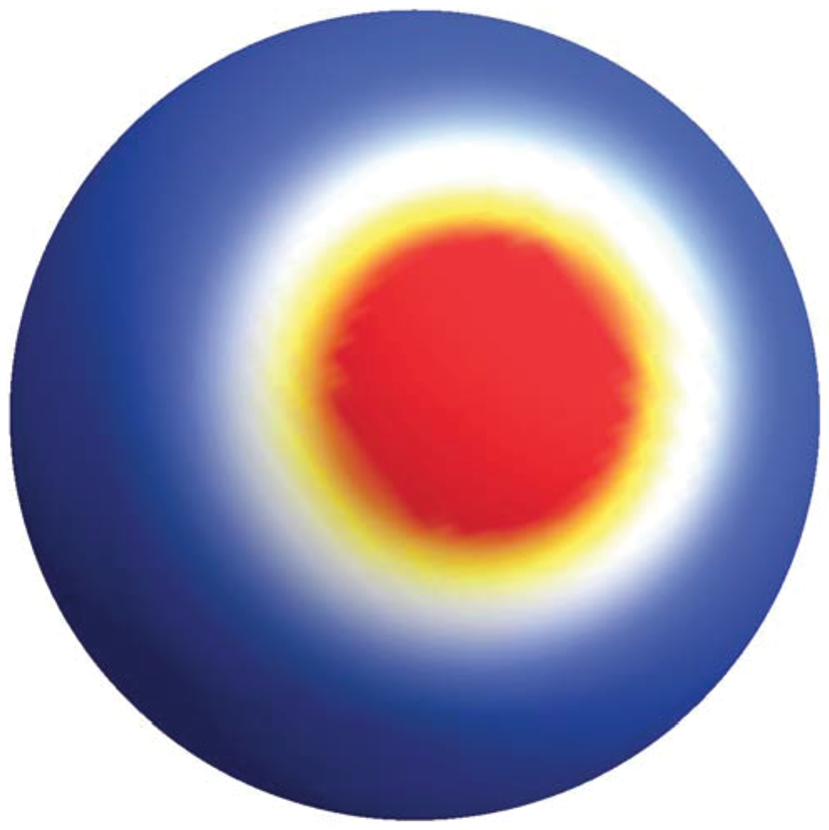}}
\scalebox{0.8}{\includegraphics{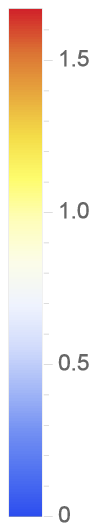}}
\scalebox{0.3}{\includegraphics{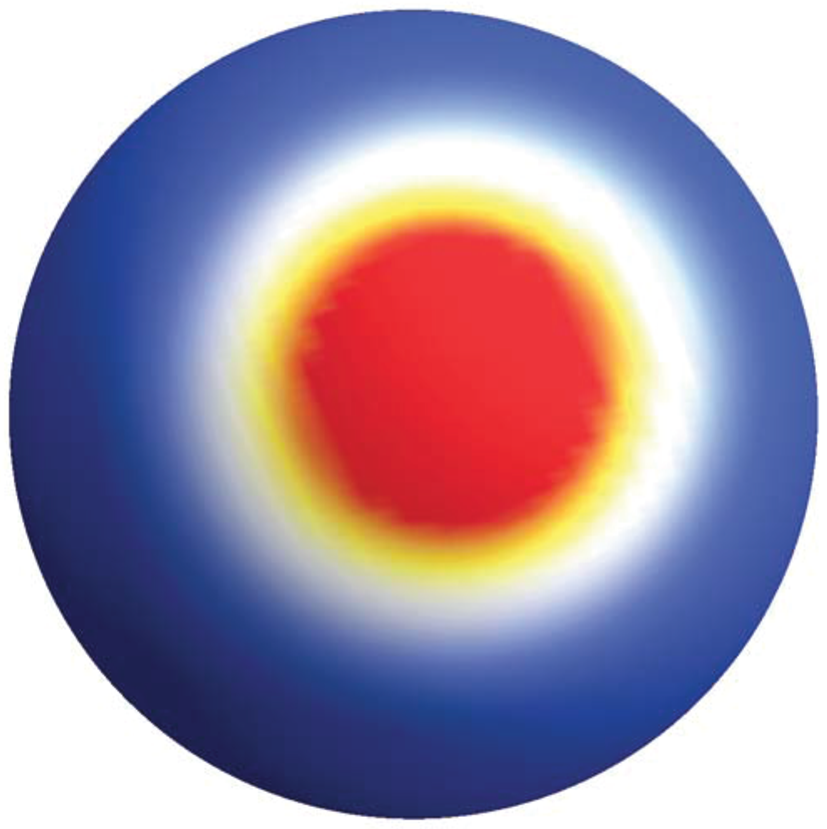}}
\scalebox{0.8}{\includegraphics{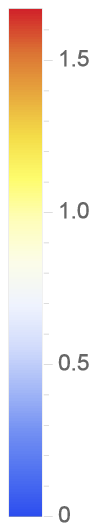}}
\scalebox{0.3}{\includegraphics{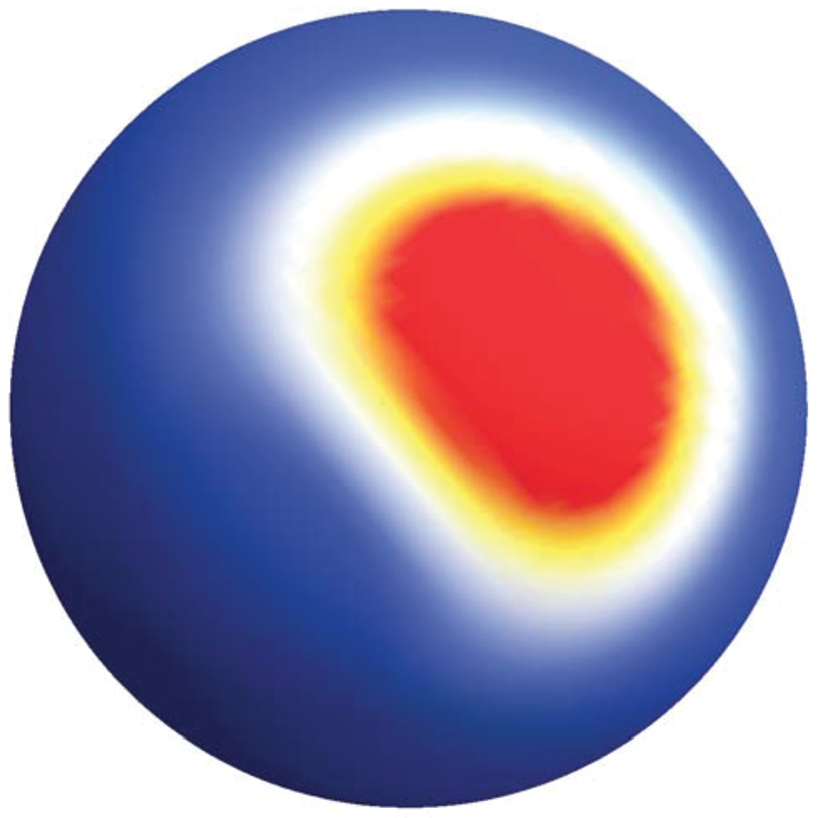}}
\scalebox{0.8}{\includegraphics{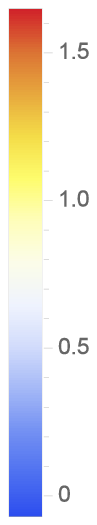}}\\
\scalebox{0.3}{\includegraphics{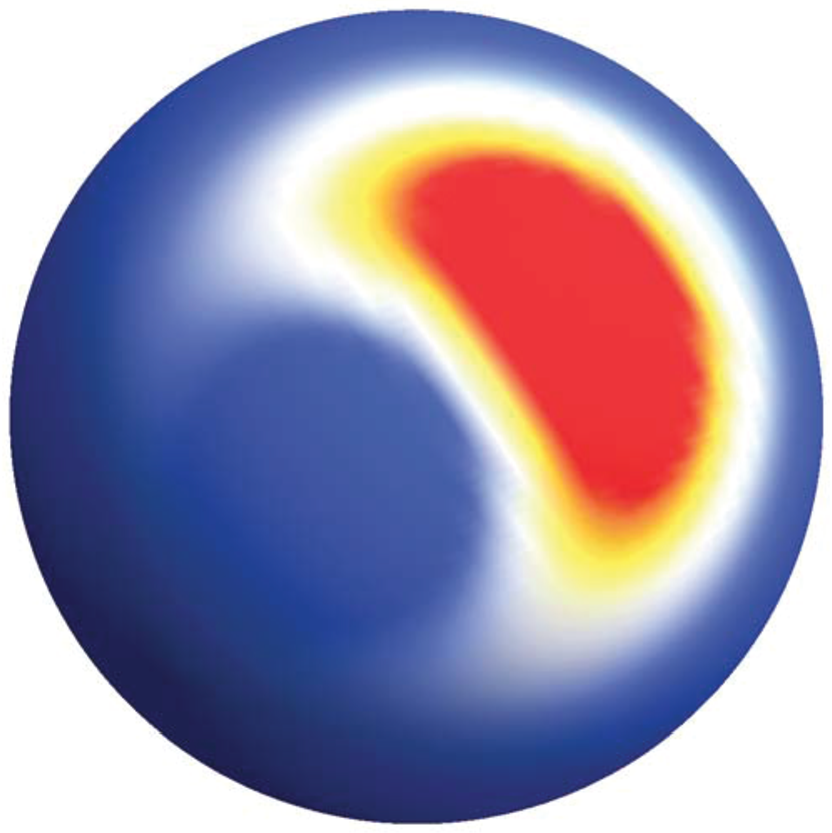}}
\scalebox{0.8}{\includegraphics{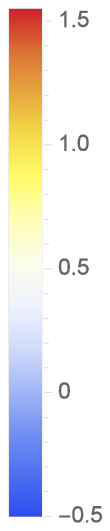}}
\scalebox{0.3}{\includegraphics{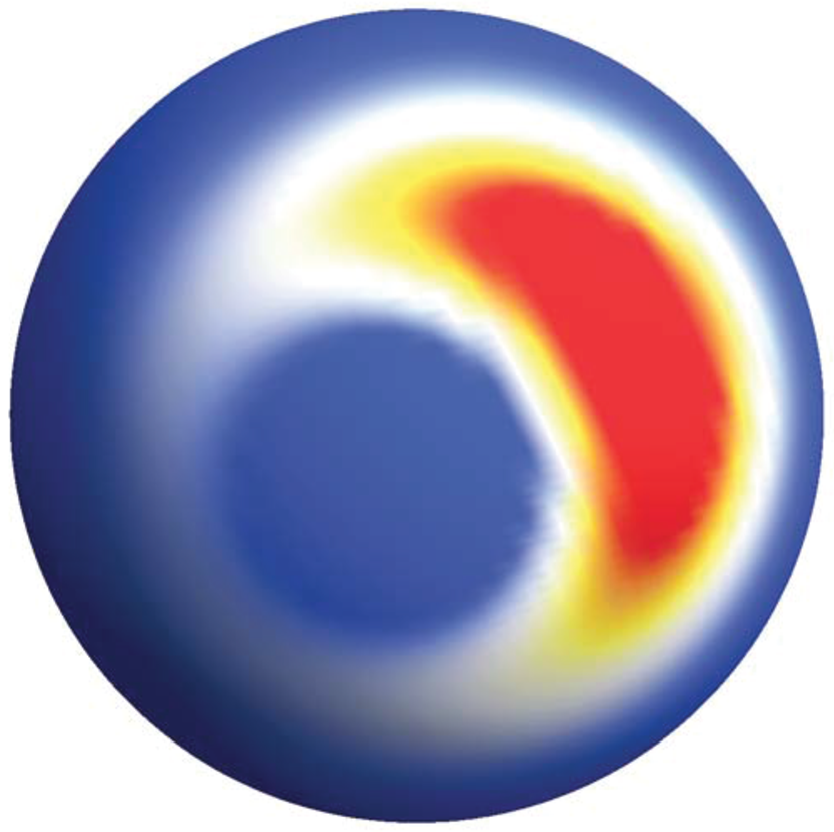}}
\scalebox{0.8}{\includegraphics{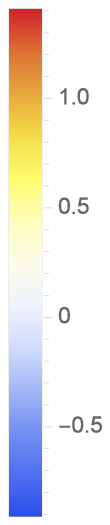}}
\scalebox{0.3}{\includegraphics{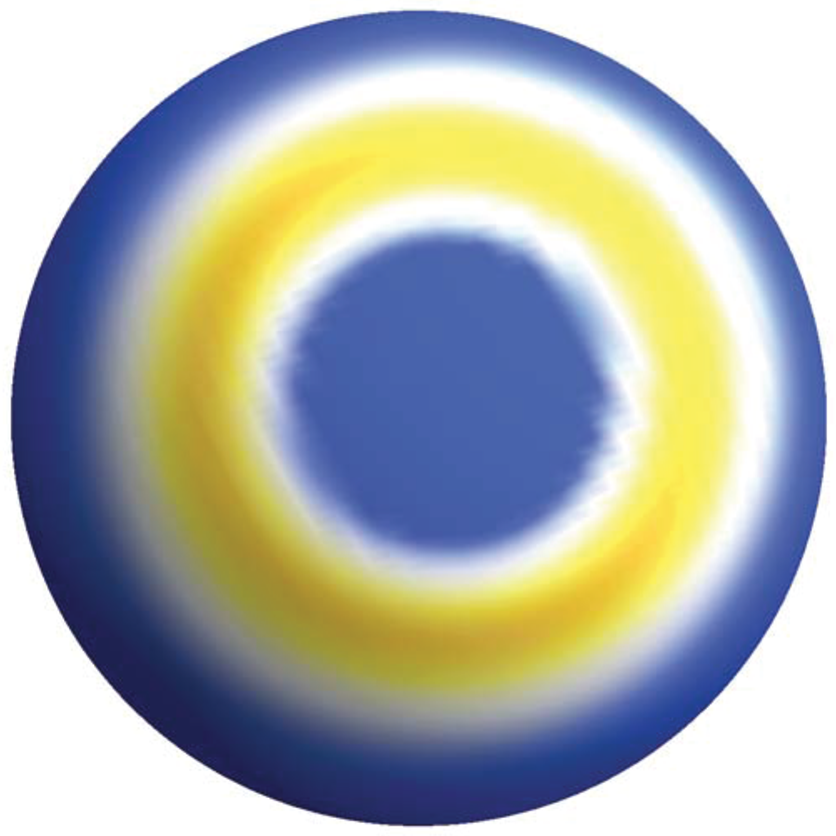}}
\scalebox{0.8}{\includegraphics{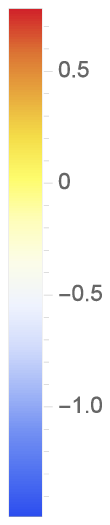}}
\caption{\label{Fig2} (Color online) The Wigner function for the initial atomic coherent state $|\frac{\pi}{2},0\rangle$ and the atomic cat state on the Bloch sphere for different $\gamma$ when $\Omega=\pi/100$ and $j=5$. The top left figure is for the atomic coherent state $|\frac{\pi}{2},0\rangle$, the top middle figure, the top right figure, the bottom left figure, the bottom middle figure, and the bottom right figure correspond to $\gamma=\pi/2, \pi/30, \pi/60, \pi/100, 0$, respectively.}
\end{center}
\end{figure}

We now demonstrate the importance of the weak value amplification and postselections in discriminating two coherent states lying nearby. For illustration purpose we choose the small phase shift $\Omega=\pi/100=1.8^{\circ}$ and the total number of the atoms $N=10$ ($j=5$). The Wigner function for the atomic cat state on the Bloch sphere for several different post-selected values of $\gamma$ is plotted in Fig.~\ref{Fig2}. We notice that the post-selected state for $\gamma=\pi/2$ can be hardly distinguished from the initial state. The post selection is useful here. We show the Wigner functions of several post-selected states by choosing $\gamma$ values such that the pre-selected and post-selected polarization states are nearly orthogonal. For decreasing values of $\gamma$ one sees more and more negative regions in the Wigner function signifying that the post-selected state becomes more and more nonclassical. The Wigner function can be determined by tomographic reconstruction \cite{Agarwal98,Auccaise,Signoles}.

\section{Amplification of the phase shift of the post-selected atomic cat state}
 In this section, we show that how the small phase shift $\Omega$ in the post-selected atomic cat state induced by the weak interaction of the atomic sample with the single photon field can be measured by studying the phase distribution of the atomic cat state.
\begin{figure}[htp]
\begin{center}
\scalebox{1.2}{\includegraphics{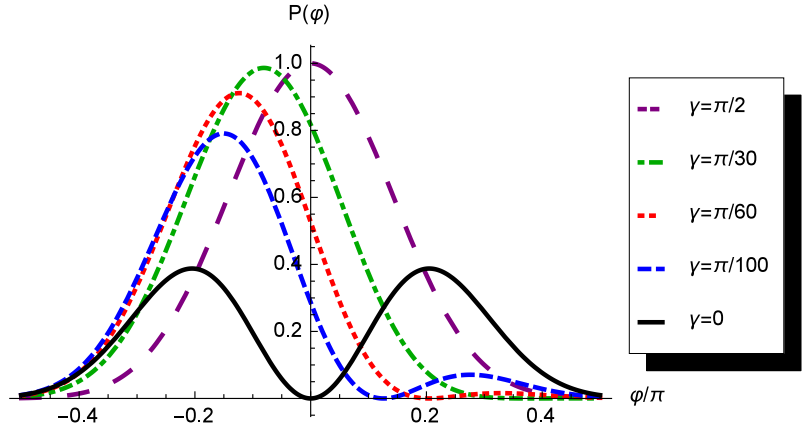}}
\caption{\label{Fig3} (Color online) The probability distribution $P(\varphi)$ as a function of $\varphi$ for different $\gamma$ when $\Omega=\pi/100$ and $j=5$. The purple long-dashed, green dotdashed, red dotted, blue short-dashed, black solid curves correspond to $\gamma=\pi/2, \pi/30, \pi/60, \pi/100, 0$, respectively.}
\end{center}
\end{figure}

\begin{figure}[htp]
\begin{center}
\scalebox{0.8}{\includegraphics{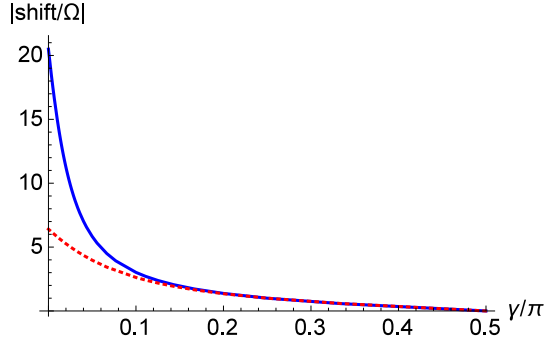}}
\caption{\label{Fig4} (Color online) The scaled left peak shift $|\mbox{shift}/\Omega|$ from $\varphi=0$ as a function of $\gamma$ for $\Omega=\pi/100$. The blue solid and red dotted curves correspond to $j=5, 50$, respectively.}
\end{center}
\end{figure}

The phase distribution $P(\varphi)$ for the atomic cat state \cite{Agarwal96}, i.e., the probability distribution $P(\varphi)$ of finding the system in the state $|\frac{\pi}{2},\varphi\rangle$, is given by
\begin{eqnarray}\label{21}
P(\varphi)&=&|\langle \frac{\pi}{2},\varphi|\Psi_{cat}\rangle|^2\nonumber\\
&=&\frac{1}{\mathcal{N}^2}\left(\frac{1}{2}\right)^{4j+1} \Bigg|\sum^{+j}_{m=-j}\left(\begin{array}{c} 2j \\ j+m\end{array}\right)e^{i(j+m)\varphi}\sin(\gamma-\frac{\pi}{4})e^{-im\Omega}\nonumber\\
& &+\sum^{+j}_{m=-j}\left(\begin{array}{c} 2j \\ j+m\end{array}\right)e^{i(j+m)\varphi}\cos(\gamma-\frac{\pi}{4})e^{im\Omega}\Bigg|^2.
\end{eqnarray}
The figure~\ref{Fig3} shows that the phase distribution $P(\varphi)$ as a function of $\varphi$ for different values of $\gamma$ for $\Omega=\pi/100$ and $j=5$. As the value of $\gamma$ is decreased from $\pi/2$ to $0$, the phase distribution $P(\varphi)$ gradually changes from a single peak at $\varphi=0$ to two symmetric peaks with a dip at $\varphi=0$, which is due to the destructive interference between two weakly separated coherent states in the atomic cat state. The scaled left peak shift $|\mbox{shift}/\Omega|$ from $\varphi=0$ as a function of $\gamma$ is shown in Fig.~\ref{Fig4}. It is noted that the scaled peak shift $|\mbox{shift}/\Omega|$ increases with decreasing $\gamma$. When $\gamma=\pi/100$, $|\mbox{shift}/\Omega|\simeq15$, the amplification factor for the weak value $\Omega$ is about 15.
\begin{figure}[htp]
\begin{center}
\scalebox{1.2}{\includegraphics{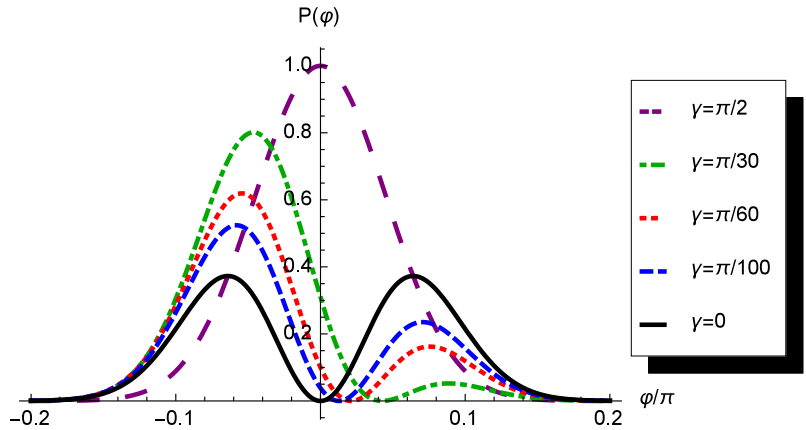}}
\caption{\label{Fig5} (Color online) The probability distribution $P(\varphi)$ as a function of $\varphi$ for different $\gamma$ when $\Omega=\pi/100$ and $j=50$.  The purple long-dashed, green dotdashed, red dotted, blue short-dashed, black solid curves correspond to $\gamma=\pi/2, \pi/30, \pi/60, \pi/100, 0$, respectively.}
\end{center}
\end{figure}

 If there is 100 atoms in the atomic sample ($j=50$), the phase distribution $P(\varphi)$ as a function of $\varphi$ for different values of $\gamma$ is shown in Fig.~\ref{Fig5}. The scaled left peak shift $|\mbox{shift}/\Omega|$ from $\varphi=0$ as a function of $\gamma$ is shown in Fig.~\ref{Fig4}. As the value of $\gamma$ is decreased, the scaled left peak shift $|\mbox{shift}/\Omega|$ increases. When $\gamma=\pi/100$, $|\mbox{shift}/\Omega|\simeq5.8$, the amplification factor for the weak value $\Omega$ is about 5.8. Note that the amplification factor for the weak value $\Omega$ decreases with increasing the total number $N$ of the atoms. This is because the modulus squared of the overlap of the two atomic coherent states $|\frac{\pi}{2},\Omega\rangle$ and $|\frac{\pi}{2},-\Omega\rangle$ decreases when the total number $N$ of the atoms becomes larger, which is shown below. Using Eq. (\ref{10}), one gets
\begin{equation}\label{22}
|\langle \frac{\pi}{2},\Omega|\frac{\pi}{2},-\Omega\rangle|^2=\cos^{4j}\Omega,
\end{equation}
when $\Omega=\pi/100$, one has
\begin{eqnarray}\label{23}
|\langle \frac{\pi}{2},\Omega|\frac{\pi}{2},-\Omega\rangle|^2=\left\{\begin{array}{cc}
0.990 & (N=10,j=5),\\
0.906 & (N=100,j=50).\\
\end{array}\right.
\end{eqnarray}
And $\frac{\cos^{20}\big(\frac{\pi}{100}\big)}{\cos^{200}\big(\frac{\pi}{100}\big)}=1.092$.

We now find the observable whose weak value is being displayed in Figs.~\ref{Fig3}--\ref{Fig5}. To see this we write the numerator in (\ref{21}) as $|D(\varphi)|^2$ where
\begin{eqnarray}
D(\varphi)&=&\sin(\gamma-\frac{\pi}{4})f(\varphi-\Omega)+\cos(\gamma-\frac{\pi}{4})f(\varphi+\Omega),\label{24}\\
f(\varphi)&=&\sum^{+j}_{m=-j}\left(\begin{array}{c} 2j \\ j+m\end{array}\right)e^{im\varphi}.\label{25}
\end{eqnarray}
To simplify (\ref{24}) for small enough $\Omega$, we do a Taylor series expansion
\begin{eqnarray}
D(\varphi)&\approx&[\sin(\gamma-\frac{\pi}{4})+\cos(\gamma-\frac{\pi}{4})][f(\varphi)+\Omega f'(\varphi)A],\label{26}\\
A&=&\frac{-\sin(\gamma-\frac{\pi}{4})+\cos(\gamma-\frac{\pi}{4})}{\sin(\gamma-\frac{\pi}{4})+\cos(\gamma-\frac{\pi}{4})}=\cot(\gamma).\label{27}
\end{eqnarray}
Thus we can rewrite (\ref{26}) as
\begin{eqnarray}
D(\varphi)\approx f(\varphi+A\Omega)[\sin(\gamma-\frac{\pi}{4})+\cos(\gamma-\frac{\pi}{4})].\label{28}
\end{eqnarray}
The expansion (\ref{28}) is valid unless $\gamma$ is very small. Thus the shift of the distribution is $-A\Omega$ where $A$ is the weak value of the observable defined by the Stokes operator for the photon field
\begin{eqnarray}
S&=&|1_{+},0_{-}\rangle\langle 1_{+},0_{-}|-|0_{+},1_{-}\rangle\langle 0_{+},1_{-}|,\label{29}\\
A&=&-\frac{\langle \Psi_{f}|S|\Psi_{ph}\rangle}{\langle\Psi_{f}|\Psi_{ph}\rangle},\label{30}
\end{eqnarray}
where $|\Psi_{f}\rangle$ and $|\Psi_{ph}\rangle$ are defined respectively by Eqs. (\ref{11}) and (\ref{13}). Here $|\Psi_{f}\rangle$ is the initial state of the photon and $|\Psi_{ph}\rangle$ is the post-selected state of the photon.

Before concluding this section we present a method of studying the post-selected cat state (\ref{15}). The best way to study this cat state is via the distributions of the spin variables. Note that the basic interaction (\ref{4}) produced the heralded cat state (\ref{15}) by using a single photon. Now imagine that after heralding a second classical field with linear polarization is sent through the ensemble, then the polarization of the classical field will be rotated. This rotation of polarization will depend on the quantum variable $J_{z}$. Thus the studies in the fluctuations of the polarization of the classical field applied after initial heralding will measure the quantum characteristics of the cat state. Such techniques were initially pioneered by Julsgaard et. al. \cite{Julsgaard} and more recently applied to heralded cat states by McConnell et. al. \cite{McConnellNat}. Further information on the cat state (\ref{15}) can be obtained by using a rf field so as to work in a different basis \cite{Agarwal98,Auccaise,Signoles}. This way one can do a complete tomography of the state.

\section{Quantum Fisher information and the weak value amplification of $\Omega$}
The precision in estimating the parameter $\Omega$ is limited by the quantum Cramer-Rao bound, which gives the minimum achievable variance $\Delta^2\Omega$, depending on the quantum Fisher information \cite{Caves, Yamamoto}.
In this section, we examine the question of the estimation of the parameter $\Omega$ via the quantum Fisher information. The utility of the weak value amplification in comparison of the standard metrological protocols has been questioned \cite{Jordan, Zhang13, Caves, Yamamoto}. However, Alves et. al. \cite{Alves} have shown that for most preselected states, full information on the coupling constant (analog of $\Omega$) can be obtained from the meter data while for a small fraction of the pre-selected states, the full information must be obtained from the post-selection statistics. In a comprehensive paper Jordan et. al. \cite{Jordanprx} have shown theoretically that in presence of many different sources of noise, the weak value amplification outperforms the standard metrological protocol. These theoretical results have been given experimental support \cite{Torres, Brunner,Jordanexp}. The quantum Fisher information $I_{at-f}$ in the state (\ref{12}) defined by
\begin{eqnarray}\label{31}
I_{at-f}&=&4\left[\left(\frac{\mbox{d}\langle \Psi_{at-f}|}{\mbox{d} \Omega}\right)\left(\frac{\mbox{d}|\Psi_{at-f}\rangle}{\mbox{d} \Omega}\right)-\left|\langle\Psi_{at-f}|\left(\frac{\mbox{d}|\Psi_{at-f}\rangle}{\mbox{d} \Omega}\right)\right|^2\right]
\end{eqnarray}
is given by
\begin{equation}\label{32}
I_{at-f}=N,\quad \theta=\frac{\pi}{2},\quad \phi=0,\quad  c_{\pm}=\frac{1}{\sqrt{2}}.
\end{equation}
The quantum Fisher information in the post-selected state (\ref{15}) is given by
\begin{eqnarray}\label{33}
I&=&4\left[\left(\frac{\mbox{d}\langle \Psi_{cat}|}{\mbox{d} \Omega}\right)\left(\frac{\mbox{d}|\Psi_{cat}\rangle}{\mbox{d} \Omega}\right)-\left|\langle\Psi_{cat}|\left(\frac{\mbox{d}|\Psi_{cat}\rangle}{\mbox{d} \Omega}\right)\right|^2\right]\nonumber\\
&=&\frac{0.5N}{p}\Big[1+\cos(2\gamma)\cos^{N-2}\Omega(1-N\sin^{2}\Omega)\nonumber\\
& &-\frac{0.5N}{p}\cos^{2}(2\gamma)\cos^{2N-2}\Omega\sin^{2}\Omega\Big],
\end{eqnarray}
and
\begin{eqnarray}\label{34}
p&=&\frac{1}{2}[1-\cos(2\gamma)\cos^{N}\Omega],
\end{eqnarray}
where $p$ is the probability of success.
The classical Fisher information in the post-selected process is defined by
\begin{eqnarray}\label{35}
F_{p}=\frac{1}{p}\left(\frac{dp}{d\Omega}\right)^2+\frac{1}{1-p}\left(\frac{d(1-p)}{d\Omega}\right)^2.
\end{eqnarray}
We find
\begin{eqnarray}\label{36}
F_{p}&=&\frac{N^2\cos^{2}(2\gamma)\cos^{2N-2}\Omega\sin^2\Omega}{1-\cos^{2}(2\gamma)\cos^{2N}\Omega}.
\end{eqnarray}

\begin{figure}[!h]
\begin{center}
\scalebox{1}{\includegraphics{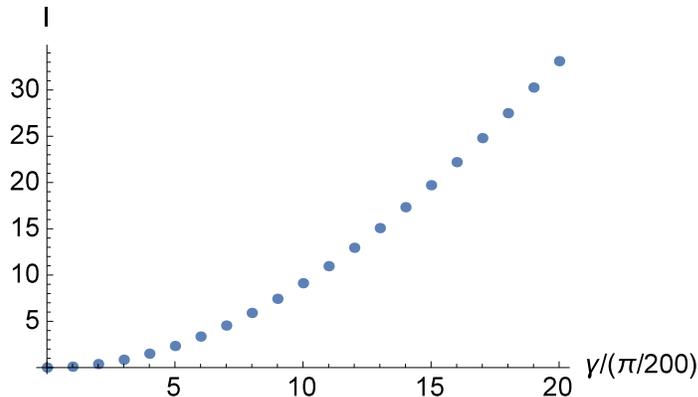}} \caption{\label{Fig6} (Color online) The quantum Fisher information $I$ of the atomic cat state as a function of $\gamma$ for $\Omega=\pi/100$ and $j=50$.}
\end{center}
\end{figure}

\begin{figure}[!h]
\begin{center}
\scalebox{1}{\includegraphics{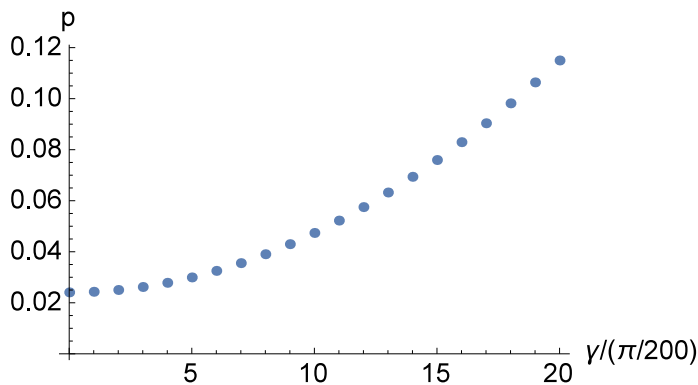}} \caption{\label{Fig7} (Color online) The probability $p$ of the successful post selection as a function of $\gamma$ for $\Omega=\pi/100$ and $j=50$.}
\end{center}
\end{figure}

\begin{figure}[!h]
\begin{center}
\scalebox{1}{\includegraphics{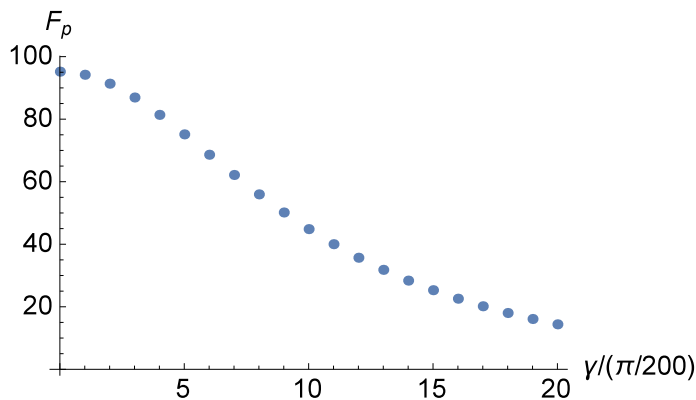}} \caption{\label{Fig8} (Color online) The classical Fisher information $F_{p}$ in the post-selected process as a function of $\gamma$ for $\Omega=\pi/100$ and $j=50$.}
\end{center}
\end{figure}
In Figs. \ref{6}-\ref{8}, we show the quantum Fisher information $I$ of the atomic cat state, the probability $p$ of the successful post selection, and the classical Fisher information $F_{p}$ in the post-selected process as a function of $\gamma$ for $\Omega=\pi/100$ and $j=50$.
 First of all, we note that although $pI\ll I_{at-f}=N$ (Eq. (\ref{32})), as expected from the analysis of Refs. \cite{Caves, Yamamoto}. On the other hand, the missing information can be obtained from the study of both $I$ and $F_{p}$ as evidenced from Figs. \ref{6}-\ref{8}. This has been emphasized in Refs. \cite{Jordanprx,Jordanexp,Alves,Hofmann}, see especially Fig. 2 in Ref. \cite{Alves}.

\section{Conclusions}
We have demonstrated how the nonclassical interference between two slightly separated atomic coherent states in a post-selected atomic cat state can be observed via making a weak value amplification on a system containing $N$ two-level atoms weakly coupled to a single photon field. We find that the negative region of the Wigner distribution of the atomic cat state is increased when the pre-selected and post-selected states of the single photon field are closer to orthogonal. We show that the weak value amplification can lead to peak shift in
the phase distribution of the atomic cat state that can be 15
times [for 10 atoms] the phase shift in the atomic cat state when the initial and final states of the signal photon field are nearly orthogonal. We have presented an experimental scheme to do tomography of the post-selected cat state. We have also discussed different aspects of the quantum and the classical Fisher information for the states of the previous sections.

\section*{Acknowledgments}
This work was supported by the Singapore National Research Foundation under NRF Grant No. NRF-NRFF 2011-07 and the Oklahoma State University. GSA thanks the hospitality of the Director, Tata Institute of Fundamental Research where this work was completed. He also thanks A. Jordan for discussions on weak values.

\end{document}